\documentclass[aps,prd,tightenlines,preprint,nofootinbib,a4paper]{revtex4}
\usepackage{epsfig}

\begin{document}

\title{Rephasing invariance and neutrino mixing}

\author{S. H. Chiu\footnote{schiu@mail.cgu.edu.tw}}
\affiliation{Physics Group, CGE, Chang Gung University, 
Kwei-Shan 333, Taiwan}

\author{T. K. Kuo\footnote{tkkuo@purdue.edu}}
\affiliation{Department of Physics, Purdue University, West Lafayette, IN 47907, USA}


\begin{abstract}

A rephasing invariant parametrization is introduced for three flavor neutrino mixing.
For neutrino propagation in matter, these parameters are shown
to obey evolution equations as functions of the induced neutrino mass.
These equations are found to preserve (approximately) some characteristic
features of the mixing matrix, resulting in solutions
which exhibit striking patterns as the induced mass varies. 
The approximate solutions are compared to numerical integrations
and found to be quite accurate.

\end{abstract}

\pacs{14.60.Pq, 14.60.Lm, 13.15.+g}

\maketitle


\section{Introduction}

Flavor mixing plays a central role in the physics of flavors.
For quarks, the CKM ($V_{CKM}$) matrix has stood the test of time and is found
to be sufficient in describing all of the relevant physics.
Similarly, the PMNS ($V_{\nu}$) matrix has been used to analyze neutrino 
oscillation with no known discrepancies.
Mathematically, both matrices belong to elements of $U(3)$, the $3 \times 3$
unitary matrices.  Physically, for quarks, since the phases of individual quark
states are unobservable, the rephasing transformation,
$V_{CKM} \rightarrow PV_{CKM}P'$, where $P$ and $P'$ are arbitrary phase matrices,
leaves the physics unchanged.  Thus, only the rephasing invariant part of $V_{CKM}$
is physical.  For $V_{\nu}$, while the charged lepton phases are unobservable,
for Majorana neutrinos actually there are two observable, 
CP-violating, phases \cite{ref1}.  
However, as long as one restricts oneself to lepton number conserving processes, 
such as in neutrino oscillations, these phases also become unphysical so that
the physical $V_{\nu}$ is again rephasing invariant.

Related to the rephasing invariance of mixing matrices is their parametrization.
While it may appear that the choice of parametrization is not important, since, 
at the end of the day, the physical quantities must be grouped into
rephasing invariant combinations.  However, when one deals with
a situation where relations between parameters are considered, 
a particular choice may be advantageous over others.
For instance, when the mixing depends on the energy scale,
as in the RGE for mass matrices, we have a set of evolution equations
relating parameters at neighboring scales.  Another example deals with
neutrino mixing in matter.  Here, the mixing depends on the density of matter
and neutrino energy, contributing to an induced neutrino mass.
One can establish the relation between parameters for neighboring densities,
resulting in a set of evolution equations as a function of the neutrino effective
mass.  They are very similar to those of the energy scale as described
by a set of RGE.  It turns out that, in both cases, the use of explicitly
rephasing invariant parameters simplifies the evolution equations.

In the following we will derive a set of evolution equations,
as a function of the effective mass of neutrinos, for neutrino parameters in matter.
These equations are based on the use of rephasing invariant parameters
developed earlier.  We find that they have simple, analytic, albeit approximate,
solutions.  It is interesting that the parameters in matter preserve a number of
salient features of those in vacuum, resulting in a
matter-dependent PMNS matrix that can be grasped at a glance.


The paper is organized as follows.
Section II is a brief summary of the rephasing invariant parametrization that is adopted
in this work. In stead of directly solving the eigenvalue problem,
we derive in Section III the evolution equations for the neutrino mixing parameters and masses 
from the effective Hamiltonian in matter.  Certain well-known invariants are also derived 
using the symmetric properties of the equations.  Section IV is devoted first to
solving the two-flavor problem using this rephasing invariant formulation, and then
the three-flavor case.  Making use of the known properties of measured neutrino parameters,
analytic, approximate, solutions are obtained.  In Section V, the accuracy of the solutions
are confirmed by comparison with numerical integration of the equations.  
Section VI is the summary.  In appendix A, we also derive the neutrino transition probabilities
in matter using the adopted rephasing invariant parametrization.



\section{Rephasing invariant parametrization}

In this section, we briefly summarize the rephasing invariant parametrization
introduced earlier for quark mixing \cite{Kuo:05}, and will now be adopted for
neutrino mixing, valid for lepton number conserving processes.

For the PMNS matrix ($V$), without loss of generality, we can
impose the condition $\mbox{det}V=+1$.
There are then a set of rephasing invariants
\begin{equation}\label{eq:g}
\Gamma_{ijk}=V_{1i}V_{2j}V_{3k}=R_{ijk}-iJ,
\end{equation}
where their common imaginary part can be identified with
the Jarlskog invariant $J$ \cite{Jar:85}.  
Their real parts are defined as
\begin{equation}
(R_{123},R_{231},R_{312};R_{132},R_{213},R_{321})
=(x_{1},x_{2},x_{3};y_{1},y_{2},y_{3}).
\end{equation}
These variables are bounded by $\pm 1$: 
$-1 \leq (x_{i},y_{j}) \leq +1$,
with $y_{j} \leq x_{i}$ for any ($i,j$). They satisfy two constraints
\begin{eqnarray}\label{cons}
\mbox{det}V=(x_{1}+x_{2}+x_{3})-(y_{1}+y_{2}+y_{3})=1, \\
(x_{1}x_{2}+x_{2}x_{3}+x_{3}x_{1})-(y_{1}y_{2}+y_{2}y_{3}+y_{3}y_{1})=0.
\end{eqnarray}
In addition, it is found that 
\begin{equation}\label{eq:J}
J^{2}=x_{1}x_{2}x_{3}-y_{1}y_{2}y_{3}.
\end{equation}

The $(x,y)$ parameters are related to $|V_{ij}|^{2}$ by
\begin{equation}\label{eq:w}
 W = \left[|V_{ij}|^{2}\right]
   = \left(\begin{array}{ccc}
                    x_{1}-y_{1} & x_{2}-y_{2}   &  x_{3}-y_{3} \\
                     x_{3}-y_{2} & x_{1}-y_{3}  & x_{2}-y_{1} \\
                    x_{2}-y_{3}  &   x_{3}-y_{1}    & x_{1}-y_{2} \\
                    \end{array}\right).  
\end{equation}
One can readily obtain the parameters $(x,y)$ from $W$ by computing its cofactors,
which form the matrix $w$ with $w^{T}W=(\mbox{det}W)I$, and is given by
\begin{equation}\label{eq:co}
 w = \left(\begin{array}{ccc}
                    x_{1}+y_{1} & x_{2}+y_{2}   &  x_{3}+y_{3} \\
                     x_{3}+y_{2} & x_{1}+y_{3}  & x_{2}+y_{1} \\
                    x_{2}+y_{3}  &   x_{3}+y_{1}    & x_{1}+y_{2} \\
                    \end{array}\right).     
\end{equation}
Eqs.~(\ref{eq:w}) and ~(\ref{eq:co}) establish the close relationship between
the two rephasing invariant parametrizations $(x,y)$ and $|V_{ij}|^{2}$.
Besides the obvious difference in the number of constraints 
(two for $(x,y)$ and five for $|V_{ij}|^{2}$), the set $(x,y)$ has built-in
symmetry amongst the three states considered, which, as we will see,
helps to make the evolution equations simpler. 

For the PMNS matrix in vacuum, its elements squared are well-approximated by
\begin{equation}\label{w0}
W_{0} = \left(\begin{array}{ccc}
  \frac{2(1-\epsilon^{2})}{3}-2\eta & \frac{1-\epsilon^{2}}{3}+2\eta    &  \epsilon^{2} \\
 \frac{1+2\epsilon^{2}-\xi}{6}+\beta+\eta & 
     \frac{2+\epsilon^{2}-2\xi}{6}-\beta-\eta & 
     \frac{1-\epsilon^{2}+\xi}{2} \\
 \frac{1+2\epsilon^{2}+\xi}{6}-\beta+\eta  & 
\frac{2+\epsilon^{2}+2\xi}{6}+\beta-\eta  & 
    \frac{1-\epsilon^{2}-\xi}{2} \\
                    \end{array}\right),
\end{equation}    
with $(\epsilon, \eta, \beta, \xi) \ll 1$.  
$W_{0}$ reduces to the tri-bimaximal \cite{tribi} 
matrix when $\epsilon=\eta=\beta=\xi=0$.
If we allow the parameters $(\epsilon, \eta, \beta, \xi)$ to take on arbitrary values,
the matrix above can be used as a general parametrization of the mixing matrix.
Also, it is related to the familiar ``standard parametrization" \cite{data} 
by $S^{2}_{13}=\epsilon^{2}$, $S^{2}_{12}=\frac{1}{3}+\frac{2\eta}{1-\epsilon^{2}}$, 
$S^{2}_{23}=\frac{1}{2}+\frac{1}{2}\frac{\xi}{1-\epsilon^{2}}$,
and $2\beta=(S^{2}_{23}-C^{2}_{23})[-\frac{2}{3}C^{2}_{13}+C^{2}_{12}-S^{2}_{12}S^{2}_{13}]+
4S_{12}C_{12}S_{13}C_{23}S_{23}\cos\phi$, so that, if $(\epsilon, \eta, \xi) \ll1$,
$\beta \simeq \frac{\sqrt{2}}{3}C_{\phi}S_{13}$.

The matrix $W_{0}$ in Eq.~(\ref{w0}) exhibit several interesting features.
When $\epsilon=\eta=\beta=\xi=0$, we find
$x_{10}= 1/3$, $x_{20}= 1/6$, $x_{30}= 0$,
and $x_{i0}+y_{i0}= 0$ $(i=1,2,3)$.
The conditions $x_{30}=y_{30}=0$ come from $W_{13}=0$ (so also $V_{13}=0$).
The conditions $x_{i0}+y_{i0}=0$ are equivalent to 
$W_{2i}=W_{3i}$ \cite{Wij}.
From known experimental bounds, for non-vanishing $(\epsilon,\eta,\beta,\xi)$,
these conditions are valid to $\mathcal{O}(10^{-2})$.


\section{Evolution of neutrino mixing parameters}

It is well-established that neutrino mixing is modified 
by the presence of matter \cite{MSW}.  
Their effect has been used in the analyses of solar neutrinos,
and is expected to impact those of the supernova neutrinos,
when and if they become available.
Closer to home, there is a plethora of long baseline experiments
either in operation or in the planning stage. 
For these studies, it is essential to include the matter effects
in order to understand neutrino mixing at the fundamental level.

In the literature, effort has been devoted to solving problems
along this line (see, $e.g.$, \cite{group}). 
However, the process involves 
the complication of the cubic eigenvalue problems, 
and the results are usually far from transparent for a clear
extraction of the physical implications.

In this work we study this problem from another angle. 
It is well-known that, when neutrinos propagate through matter,
the latter contributes an induced mass to the neutrinos.
Similar to the case of RGE, we may write down, as a function of the induced mass,
a set of evolution equations for the neutrino parameters. It turns out that,
with the initial conditions given by $W_{0}$ in Eq.~(\ref{w0}), we can find simple, 
approximate, solutions to these equations, as we will detail in Sec. IV.
These results were summarized in a previous publication \cite{CKL}.

To derive these equations, we start from the effective Hamiltonian for
neutrino propagation in matter 
\begin{equation}
H_{eff}=H/2E,
\end{equation}
where $H$ is given, in the flavor basis, by
\begin{equation}\label{H}
H=\left[ V_{0}
                    \left(\begin{array}{ccc}
                    m_{1}^{2} &     &   \\
                              & m_{2}^{2} &  \\
                              &       & m_{3}^{2}  \\
                    \end{array}\right)
                    V_{0}^{\dag} + 
                         \left(\begin{array}{ccc}
                    A &     &   \\
                              & 0 &  \\
                              &       & 0  \\
                    \end{array}\right)\right],
\end{equation}
where $m_{1}$, $m_{2}$, and $m_{3}$ 
are the neutrino masses in vacuum, $V_{0}$ is the mixing matrix in vacuum,
$E$ is the neutrino energy, and the induced mass 
$A=\sqrt{2}G_{F}n_{e}E$.

The matrix $H$ can be diagonalized, 
\begin{equation}
H=VDV^{\dag}=V \left(\begin{array}{ccc}
                    D_{1} &     &   \\
                              &D_{2} &  \\
                              &       & D_{3}  \\
                    \end{array}\right) V^{\dag},
\end{equation}
where $D_{i}=M_{i}^{2}$ is the squared mass in matter.
To study how the elements of $V$ evolve in matter, one may start with
$dH/dA$, which leads to
\begin{equation}\label{eq:matrix}
V^{\dag}\frac{d}{dA}[VDV^{\dag}]V 
=\left(\begin{array}{ccc}
                    |V_{11}|^{2} & V_{12}V_{11}^{*}   &  V_{13}V_{11}^{*} \\
                     V_{11}V_{12}^{*} & |V_{12}|^{2}  & V_{13}V_{12}^{*} \\
                    V_{11}V_{13}^{*}  &   V_{12}V_{13}^{*}    & |V_{13}|^{2} \\
                    \end{array}\right).
\end{equation}
Taking the diagonal terms of Eq.~(\ref{eq:matrix}), we find 
\begin{equation}\label{eq:di}
\frac{dD_{i}}{dA}= |V_{1i}|^{2}=x_{i}-y_{i}, \hspace{.2in} (i=1,2,3).
\end{equation}

The off-diagonal terms yield
\begin{equation}\label{eq:VV}
[(\frac{dV^{\dag}}{dA})V]_{ik}=\frac{V^{*}_{1i}V_{1k}}{D_{i}-D_{k}}, \hspace{.2in} (i \neq k).
\end{equation}
The diagonal elements $[(dV^{\dag}/dA)V]_{ii}$ are not constrained by Eq.~(\ref{eq:matrix}).
Fortunately, it is rephasing dependent \cite{Chiu:09}, and we can set it to
vanish by a proper choice of the phase.  This means that, when we
multiply Eq.~(\ref{eq:VV}) by $(V^{\dag})_{kj}$, and sum over $k \neq i$ on the
right hand side, we may sum over all $k$-values on the left. The result is
\begin{equation}\label{eq:dV}
\frac{dV_{ij}}{dA}
=\sum_{k\neq j} \frac{V_{ik}V_{1j}}{D_{j}-D_{k}} V_{1k}^{*}.
\end{equation} 
Note that the dependence is only on the mass differences, $(D_{j}-D_{k})$,
in accordance with the invariance
of $V_{ij}$ if $H \rightarrow H+\mbox{constant}$.

While Eq.~(\ref{eq:dV}) is valid only with a particular choice of phase,
this rephasing ambiguity is removed if one uses it to compute
rephasing invariant quantities, $e.g.$,
\begin{equation}
\frac{d\Gamma_{123}}{dA} =\frac{d}{dA}(V_{11}V_{22}V_{33})= \frac{dx_{1}}{dA}-i\frac{dJ}{dA}.         
 \end{equation}
After some algebra, separating the real and imaginary parts,
in addition to using different $\Gamma^{'}_{ijk}s$, 
we obtain the evolution equations for all $(x_{i},y_{i})$ and
$d\ln J/dA$, which are collected in Table I.

Note that, since Eq.~(\ref{eq:dV}) can be obtained from Eq.(3.6) (Ref. \cite{Chiu:09}) in
appropriate limits, the entries in Table I can be identified with those
in Table II of Ref. \cite{Chiu:09}.
Indeed, it can be verified that 
$dx_{i}/dA=\sum[(\bar{B}_{i})_{2r}-(\bar{B}_{i})_{3r}]/(D_{s}-D_{t})$,
$dy_{j}/dA=\sum[(\bar{B}'_{j})_{2r}-(\bar{B}'_{j})_{3r}]/(D_{s}-D_{t})$,
where the sum is over cyclicly permuted $(r,s,t)=(1,2,3)$, and $\bar{B}_{i} (\bar{B}'_{j})$
are obtained from $B_{i} (B'_{j})$ in Table II of Ref. \cite{Chiu:09}
by exchanging $x_{2} \leftrightarrow x_{3}$,
since $V \leftrightarrow V^{\dag}$ under the usual conventions in going from
quarks to neutrinos.


	\begin{table*}[ttt]
  \centering
	\begin{center}
 \begin{tabular}{cccc}  
 
 \hline

       & $1/(D_{1}-D_{2})$ & $1/(D_{2}-D_{3})$  & $1/(D_{3}-D_{1})$    \\ \hline
    $dx_{1}/dA$ & $x_{1}x_{2}-2x_{1}y_{2}+y_{1}y_{2}$  & 
    $-x_{1}x_{2}+x_{1}x_{3}+y_{1}y_{2}-y_{1}y_{3}$ & 
    $-x_{1}x_{3}+2x_{1}y_{3}-y_{1}y_{3}$          \\
     $dx_{2}/dA$  & $-x_{1}x_{2}+2x_{2}y_{1}-y_{1}y_{2}$  & 
     $x_{2}x_{3}-2x_{2}y_{3}+y_{2}y_{3}$  & 
     $x_{1}x_{2}-x_{2}x_{3}-y_{1}y_{2}+y_{2}y_{3}$     \\
   $dx_{3}/dA$ & $-x_{1}x_{3}+x_{2}x_{3}+y_{1}y_{3}-y_{2}y_{3}$ & 
   $-x_{2}x_{3}+2x_{3}y_{2}-y_{2}y_{3}$ & $x_{1}x_{3}-2x_{3}y_{1}+y_{1}y_{3}$     \\                     
   $dy_{1}/dA$ & $ -x_{1}x_{2}+2x_{2}y_{1}-y_{1}y_{2}$ &
   $-x_{1}x_{2}+x_{1}x_{3}+y_{1}y_{2}-y_{1}y_{3}$ & 
   $x_{1}x_{3}-2x_{3}y_{1}+y_{1}y_{3}$    \\
   $dy_{2}/dA$ & $x_{1}x_{2}-2x_{1}y_{2}+y_{1}y_{2}$  &
   $-x_{2}x_{3}+2x_{3}y_{2}-y_{2}y_{3}$  & 
   $x_{1}x_{2}-x_{2}x_{3}-y_{1}y_{2}+y_{2}y_{3}$ \\
   $dy_{3}/dA$ &  $-x_{1}x_{3}+x_{2}x_{3}+y_{1}y_{3}-y_{2}y_{3}$  & 
   $x_{2}x_{3}-2x_{2}y_{3}+y_{2}y_{3}$ & $-x_{1}x_{3}+2x_{1}y_{3}-y_{1}y_{3}$ \\
    $d(\ln J)/dA$  &  $-x_{1}+x_{2}+y_{1}-y_{2}$  &    $-x_{2}+x_{3}+y_{2}-y_{3}$
        & $x_{1}-x_{3}-y_{1}+y_{3}$              \\
   \hline
  \end{tabular}
    \caption{$dx_{i}/dA$, $dy_{i}/dA$, and $d(\ln J)/dA$ are expressed as sums
    of terms in $1/(D_{1}-D_{2})$, $1/(D_{2}-D_{3})$, and $1/(D_{3}-D_{1})$.}
  \end{center}
 \end{table*}


The symmetric form of these equations allows us to find readily the result:
\begin{equation}\label{JD}
\frac{d}{dA}\ln[J(D_{1}-D_{2})(D_{2}-D_{3})(D_{3}-D_{1})]=0,
\end{equation}
$i.e.$, the product $[J(D_{1}-D_{2})(D_{2}-D_{3})(D_{3}-D_{1})]$ is a constant as $A$ changes,
a well known result derived with different methods \cite{HNK}.  

From Table I, we find
\begin{equation}\label{x1y1}
\frac{1}{2}\frac{d}{dA}\ln(x_{1}-y_{1})=\frac{x_{2}-y_{2}}{D_{1}-D_{2}}-
 \frac{x_{3}-y_{3}}{D_{3}-D_{1}},
\end{equation}
\begin{equation}\label{x2y2}
\frac{1}{2} \frac{d}{dA}\ln(x_{2}-y_{2})= -\frac{x_{1}-y_{1}}{D_{1}-D_{2}}+
 \frac{x_{3}-y_{3}}{D_{2}-D_{3}},
\end{equation}
\begin{equation}\label{x3y3}
\frac{1}{2} \frac{d}{dA}\ln(x_{3}-y_{3})=-\frac{x_{2}-y_{2}}{D_{2}-D_{3}}+
 \frac{x_{1}-y_{1}}{D_{3}-D_{1}}.
\end{equation}
We see that there is another ``matter invariant":
\begin{equation}
\frac{d}{dA}[\frac{J^{2}}{(x_{1}-y_{1})(x_{2}-y_{2})(x_{3}-y_{3})}]=0.
\end{equation}                 
Or, 
\begin{equation}\label{eq:in2}
J^{2}/(|V_{11}|^{2}|V_{12}|^{2}|V_{13}|^{2})=\mbox{constant}.
\end{equation}
When we use the ``standard parametrization", it is seen that 
$J^{2}/(|V_{11}|^{2}|V_{12}|^{2}|V_{13}|^{2})=S^{2}_{\phi}S^{2}_{23}C^{2}_{23}$,
$i.e.$, $S_{\phi}\sin2\theta_{23}$ is independent of $A$, 
a result obtained earlier \cite{TP}.

The evolution equations for $(x,y)$ also have a structure
akin to that of the fixed point of single variable equations.    
It can be verified that, if $x_{i}+y_{i}=0$ $(i=1,2,3)$, then 
\begin{equation}\label{sigma}
\frac{d}{dA}(x_{j}+y_{j})=0,  \hspace{.2in} j=(1,2,3).
\end{equation} 
This result is understandable since the conditions
$x_{i}+y_{i}=0$ are equivalent to $W_{2i}=W_{3i}$, which,
in turn, imply that the effective Hamiltonian $H$ has a 
$\mu-\tau$ exchange symmetry \cite{23-sym}.
This symmetry is clearly independent of $A$ in Eq.~(\ref{H}),
resulting in Eq.~(\ref{sigma}).  Note also that there are actually
only two independent constraints in $x_{i}+y_{i}=0$.
Given any two of them, say for $i=1,2$, we can use Eq.(4)
to derive $x_{3}+y_{3}=0$.  Thus, the set of evolution
equations has a ``fixed surface" (in the four-dimensional parameter space), points on the
surface defined by $x_{i}+y_{i}=0$ stay on it as $A$ varies.  


\section{Approximate Solutions}

While analytical solutions to the equations in Table I are not available, 
as we will see, given the known physical parameters, one can exploit certain
characteristic properties thereof to arrive at simple,
but fairly accurate, solutions to these equations.
Before we do that, it is instructive to first study the two flavor problem,
which can be compared to the traditional approach, since exact solutions can be
obtained in both cases.

\subsection{Two-flavor problem}

For two flavors, we have
\begin{equation}
\frac{dH}{dA}=\left(\begin{array}{cc}
                    1 & 0  \\
                    0 & 0 \\     
                    \end{array}\right),
\end{equation}
with the familiar diagonalization matrix
\begin{equation}
V=\left(\begin{array}{cc}
                    \cos\theta & \sin\theta  \\
                    -\sin\theta & \cos\theta \\     
                    \end{array}\right),
\end{equation}
so that $x=V_{11}V_{22}=\cos^{2}\theta$, $y=V_{12}V_{21}=-\sin^{2}\theta$, $x-y=1$.
The evolution equations are
\begin{eqnarray}\label{evo}
\frac{dD}{dA}&=&-(x+y), \nonumber \\
\frac{dx}{dA}&=& \frac{2xy}{D}=\frac{dy}{dA}, 
\end{eqnarray}
where $D \equiv D_{2}-D_{1}$.
It follows that 
\begin{equation}\label{alpha}
\frac{d}{dA}(xyD^{2})=0,
\end{equation}
\begin{equation}\label{beta}
\frac{d}{dA}[D (x+y)]=-(x+y)^{2}+4xy=-1.
\end{equation}
Eq.~(\ref{alpha}) is the familiar result
\begin{equation}
D^{2}\sin^{2}2\theta=D^{2}_{0} \sin^{2} 2\theta_{0}.
\end{equation}
Eq.~(\ref{beta}) gives
\begin{equation}
D (x+y)=-A+D_{0}(x+y)_{0},
\end{equation}
and thus
\begin{equation}
D^{2}=[(A-D_{0}\cos 2\theta_{0})^{2}+D^{2}_{0} \sin^{2} 2\theta_{0}],
\end{equation}
which is the well-known resonance formula with $D=\mbox{min.}$
at $A=D_{0}\cos 2\theta_{0}$.

These results  show that the use of evolution equations is equivalent to
the traditional method, that of finding directly the eigenvalues 
of the effective Hamiltonian.  We now turn to the case of three flavors. 
                    

\subsection{Three-flavor problem}

Experimentally, it is known that 
$\delta_{0}=m^{2}_{2}-m^{2}_{1} \cong 7.6 \times 10^{-5} eV^{2}$, 
$\Delta_{0}=m^{2}_{3}-m^{2}_{2} \cong 2.4 \times 10^{-3} eV^{2}$, so that
$\delta_{0}/\Delta_{0} \ll 1$ (We assume the ``normal" ordering of neutrino masses.
The ``inverted" case can be similarly treated).
Note that these values are relevant to long baseline
experiments since 
$A=\sqrt{2}G_{F}n_{e}E \sim (7.6 \times 10^{-5} eV^{2})(E/GeV)(\rho/g cm^{-3})$.

Since $\delta_{0} \ll \Delta_{0}$, we expect that the three-flavor problem
can be approximated by a pair of well separated two-flavor problems \cite{KP:89}. 
Indeed, the structure of the differential equations in Table I shows that
the variables $(x_{i},y_{i})$ evolve slowly as a function of $A$ except for
two regions, where $D_{1} \approx D_{2}$ and $D_{2} \approx D_{3}$,
corresponding to the two resonance regions.  
More precisely, let us denote by $(A_{0},A_{l},A_{i},A_{h},A_{d})$ the values of $A$
in vacuum $(A_{0}=0)$, at the lower resonance $(A_{l}, [d(D_{1}-D_{2})/dA]_{A_{l}}=0)$,
the intermediate range $(A_{i})$, the higher resonance $(A_{h}), [d(D_{2}-D_{3})/dA]_{A_{h}}=0)$,
and for dense medium $(A_{d})$.  Rapid evolution for $(x_{i},y_{i})$ only occurs for
$A\approx A_{l}$ and $A\approx A_{h}$. 
Near the lower resonance region, $(D_{2}-D_{1}) \ll (D_{3}-D_{1})$ or $(D_{3}-D_{2})$.
For the higher resonance region, $(D_{3}-D_{2}) \ll (D_{3}-D_{1})$ or $(D_{2}-D_{1})$.
Thus, for these two regions, we need only to keep terms $\propto 1/(D_{2}-D_{1})$
and $1/(D_{3}-D_{2})$, respectively.  This approximation is generally valid to
$\mathcal{O}(10^{-2})$.  We now turn to a detailed analysis.

For $0<A<A_{i}$, in the neighborhood of $A_{l}$, we keep terms $\propto 1/(D_{1}-D_{2})$
in Table I.  Let's concentrate on the variables $x_{i}-y_{i}=|V_{1i}|^{2}$ and use
Eq.~(\ref{eq:di}) and Eqs.~(\ref{x1y1})-(\ref{x3y3}).
We define
\begin{eqnarray}
X&=&x_{1}-y_{1}, \nonumber \\
Y&=&x_{2}-y_{2}, \nonumber \\
Z&=&x_{3}-y_{3}, \nonumber \\
\delta&=&D_{2}-D_{1}.
\end{eqnarray}
Then,
\begin{eqnarray}\label{eq:XZ}
\frac{dX}{dA}&=&-\frac{2XY}{\delta}=-\frac{dY}{dA}, \nonumber \\
\frac{dZ}{dA}&=&0, \nonumber \\
\frac{d\delta}{dA}&=&-(X-Y).
\end{eqnarray}
These equations are identical to those for the two flavor problem, Eq.~(\ref{evo}),
with $X \rightarrow x$, $Y \rightarrow -y$, and $\delta \rightarrow D$.
Also, in place of $x-y=1$, we have
\begin{equation}
X+Y=p_{l},
\end{equation}
where $p_{l}$ is a constant (since $d(X+Y)/dA=0$), and $p_{l}=1-Z$, $Z=|V_{13}|^{2}=\mbox{constant}$.
The solutions are
\begin{eqnarray}\label{sol-low}
XY\delta^{2}&=&X_{0}Y_{0}\delta^{2}_{0}, \nonumber \\
(X-Y)\delta&=&-p^{2}_{l} A+q_{l} \delta_{0}, \nonumber \\
q_{l}&=&X_{0}-Y_{0}.
\end{eqnarray}
Explicitly, we have
\begin{eqnarray}\label{low}
\delta^{2}&=&p^{2}_{l}A^{2}-2q_{l}\delta_{0}A+\delta_{0}^{2},\nonumber \\
X&=&\frac{1}{2}[p_{l}-(p^{2}_{l}A-q_{l}\delta_{0})/\delta], \nonumber \\
Y&=&\frac{1}{2}[p_{l}+(p^{2}_{l}A-q_{l}\delta_{0})/\delta].
\end{eqnarray}

Thus, $(\delta,X,Y)$ exhibit the classic resonance behavior, with the resonance location at
the minimum of $\delta$:
\begin{equation}
A_{l}=(\frac{q_{l}}{p_{l}^{2}})\delta_{0}.
\end{equation}
Substituting in the vacuum input values, 
$X_{0} \cong 2/3$, $Y_{0} \cong 1/3$, $A_{l} \cong \delta_{0}/3$.
The width of the resonance is
\begin{equation}
(\delta A)_{l}=(1-q^{2}_{l}/p^{2}_{l})^{1/2}(\delta_{0}/p_{l}).
\end{equation}
For the physical PMNS matrix, $(\delta A)_{l} \simeq \delta_{0}$. This means that the intermediate
$A$ value, $A_{i}$, already starts at $A \gtrsim (2-3)\delta_{0}$.  For $A \sim A_{i}$, $\delta \rightarrow A$,
$X \rightarrow 0$, $Y \rightarrow 1$, with $p_{l} \cong 1$.

Turning to the higher resonance, we define
\begin{equation}
\Delta=D_{3}-D_{2}.
\end{equation}
The evolution equations are
\begin{eqnarray}\label{eq:XY}
\frac{dX}{dA}&=&0, \nonumber \\
\frac{dY}{dA}&=&-\frac{2YZ}{\Delta}=-\frac{dZ}{dA}, \nonumber \\
\frac{d \Delta}{dA}&=&-(Y-Z),
\end{eqnarray}
with the solutions
\begin{eqnarray}
YZ\Delta^{2}&=& Y_{0}Z_{0}\Delta^{2}_{0}, \nonumber \\
(Y-Z)\Delta &=& -p^{2}_{h}A+q_{h}\Delta_{0}, 
\end{eqnarray}
or,
\begin{eqnarray}\label{high}
\Delta^{2}&=&p^{2}_{h} A^{2}-2q_{h}\Delta_{0} A+\Delta^{2}_{0}, \nonumber \\
Y&=&\frac{1}{2}[p_{h}-(p^{2}_{h} A-q_{h} \Delta_{0})/\Delta], \nonumber \\
Z&=&\frac{1}{2}[p_{h}+(p^{2}_{h} A-q_{h} \Delta_{0})/\Delta],
\end{eqnarray}
where $p_{h}=Y_{0}+Z_{0}$, $q_{h}=Y_{0}-Z_{0}$.  Here, the values $Y_{0}$ and $Z_{0}$
are taken at $A=A_{i} \gg \delta_{0}$, so that $Y_{0} \cong 1$, $Z_{0} \cong |V_{13}|^{2} \cong 0$,
from the solutions for the lower resonance.  The position of the higher resonance is at
\begin{equation}
A_{h}=(\frac{q_{h}}{p_{h}^{2}})\Delta_{0} \cong \Delta_{0}.
\end{equation}
Its width is
\begin{equation}
(\delta A)_{h}=(1-q^{2}_{h}/p^{2}_{h})^{1/2}(\Delta_{0}/p_{h}) 
\cong 2 Z_{0}\Delta_{0} \ll \Delta_{0}.
\end{equation}

The above analyses show that the two-flavor approximation yields simple solutions to the mixing
parameters $|V_{1i}|^{2}$, for all $A$ values.  However, the vacuum mixing, given by $W_{0}$ in Eq.~(\ref{w0}),
has another important feature, namely, $(W_{0})_{2i} \cong (W_{0})_{3i}$, or $x_{i0}+y_{i0} \cong 0$.
This feature, according to Eq.~(\ref{sigma}), is preserved by the evolution equations and so
$W_{2i} \cong W_{3i}$, or $x_{i}+y_{i} \cong 0$, for all $A$.  Thus, with the known solutions for $W_{1i}$
from above, all elements of $W$ are determined by unitarity.
Explicit solutions for $W$ or $(x_{i},y_{i})$ were presented in Ref. \cite{CKL}, obtained by using both approximations.


We may divide the full range of $A$ values into a low-$A$ and a high-$A$ regions.
The former covers the range from $A=0$ to a value below the higher resonance region,
while the later starts from beyond the lower resonance region and ends at $A=\infty$.
In these regions, the evolution equations are dominated by contributions from
pole terms, $1/(D_{1}-D_{2})$ for low-$A$ and $1/(D_{2}-D_{3})$ for high-$A$.
The exact demarcation between low-$A$ and high-$A$ is not important,
since in the intermediate region contributions from either pole are small, and
there can be considerable overlap between low-$A$ and high-$A$,
corresponding to the large range of $A_{i}$.

It should be noted that ``pole dominance", 
which was used to go from Eqs.~(\ref{x1y1}-\ref{x3y3}) to Eqs.~(\ref{eq:XZ})
and ~(\ref{eq:XY}), is an excellent approximation in this situation.  
This is because the terms dropped are doubly suppressed, first by the large
denominators, and then by the small numerators
($ (x_{3}-y_{3}) \ll 1$ throughout the low-$A$ region and 
$(x_{1}-y_{1}) \ll 1$ for high-$A$).
We should also emphasize that Eqs.~(\ref{eq:XZ})
and ~(\ref{eq:XY}) are derived independently
of the approximations $x_{i}+y_{i} \simeq 0$.  
While in our earlier work \cite{CKL}, they were
used to arrive at similar equations of evolution.

The ``pole dominance" approximation, less accurately, may also be used for
other variables, such as $J^{2}$.  This gives rise to relations,
which we shall dub as ``partial matter invariants", valid only for
either the low-$A$ or the high-$A$ regions.  Thus, from Eq.~(\ref{sol-low})
and using Table I, 
\begin{eqnarray}\label{pmiL}
(D_{1}-D_{2})^{2}|V_{11}|^{2}|V_{12}|^{2} &\cong& \mbox{constant}, 
\hspace{0.2in} (\mbox{low-A}) \nonumber \\
J^{2}(D_{1}-D_{2})^{2} &\cong& \mbox{constant}.
\end{eqnarray}
Similarly,
\begin{eqnarray}\label{pmiH}
(D_{2}-D_{3})^{2}|V_{12}|^{2}|V_{13}|^{2} &\cong& \mbox{constant}, 
\hspace{0.2in} (\mbox{high-A}) \nonumber \\
J^{2}(D_{2}-D_{3})^{2} &\cong& \mbox{constant}.
\end{eqnarray}
These ``partial matter invariants" are useful in understanding some detail properties
of the parameters. $E.g.$, for $A \sim A_{i}$, 
$|V_{11}|^{2}\cong (2/9)(\delta_{0}/A)^{2}$.  The behavior of $J^{2}$ is also clarified,
as we will see in the discussion on Fig. 4.

\begin{figure}[ttt]
\caption{Numerical (solid) and approximate (dot-dashed) solutions for 
(a) all $D_{3}(A)$, $D_{2}(A)$, and $D_{1}(A)$, and 
(b) the enlarged plot of $D_{2}(A)$ and $D_{1}(A)$ in $0 \leq A/\delta_{0} \leq 10$.
We have used $\delta_{0}/\Delta_{0}=1/32$.} 
\centerline{\epsfig{file=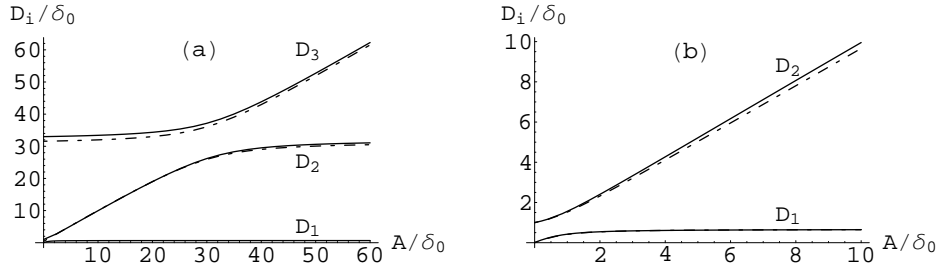,width=13 cm}}
\end{figure} 

In summary, the solution to the three flavor problem can be made simple by dividing
the full range of $A$ into a low-$A$ and a high-$A$ regions,  In the low-$A$ region,
the evolution equations are dominated by pole terms $\propto 1/(D_{1}-D_{2})$,
and the solution centers around the lower resonance.
For the high-$A$ region, correspondingly, pole terms $\propto 1/(D_{2}-D_{3})$
dominate, and the solution can be characterized by the higher resonance.
The mixing parameters change appreciably only in two regions:
1) lower resonance,
$[A_{l}-(\delta A)_{l}] \lesssim A \lesssim [A_{l}+(\delta A)_{l}]$, 
$A_{l} \cong \delta_{0}/3$, $(\delta A)_{l} \cong \delta_{0}$;
2) higher resonance,  
$[A_{h}-(\delta A)_{h}] \lesssim A \lesssim [A_{h}+(\delta A)_{h}]$, $A_{h} \cong \Delta_{0}$, 
$(\delta A)_{h} \cong 2|V_{13}|^{2}\Delta_{0} \ll \Delta_{0}$.
The solutions for $X (|V_{11}|^{2})$, $Y(|V_{12}|^{2})$, and $Z (|V_{13}|^{2})$ are:
1) $A=0$, $X_{0} \cong 2/3$, $Y_{0} \cong 1/3$, $Z_{0} \cong \epsilon^{2} \ll 1$, which are the given
vacuum values; 2) $A=A_{l}$, $X \cong Y \cong 1/2$, $Z \cong \epsilon^{2} \cong 0$;
3) $A=A_{i}$, $A_{i}$ covers roughly the range, $2\delta_{0} \lesssim A_{i} \lesssim \Delta_{0}(1-2\epsilon^{2})$,
$X \cong 0$, $Y \cong 1$, $Z \cong \epsilon^{2}$; 4) $A=A_{h}$, $X \cong 0$, $Y \cong Z \cong 1/2$;
5) $A=A_{d}$, with $A_{d} \gtrsim \Delta_{0} (1+2\epsilon^{2})$, $X\cong Y \cong 0$, $Z \cong 1$.

When we incorporate the other feature of the vacuum PMNS matrix, that $(W_{0})_{2i} \cong (W_{0})_{3i}$,
which is preserved by the evolution equations, the result is that $W_{2i} \cong W_{3i}$, for all $A$.
Given $W_{1i}$ from above, the matrix $W$ is then completely determined by unitarity.

Our results can be put together by giving the matrices $W$ at $A=(A_{0},A_{l},A_{i},A_{h},A_{d})$:
\begin{eqnarray}\label{eq:sum}
W_{0}  & \cong & \left(\begin{array}{ccc}
                    2/3 & 1/3   &  0 \\
                    1/6 & 1/3  & 1/2 \\
                   1/6  &   1/3   & 1/2 \\
                    \end{array}\right), \hspace{.15in}
    W_{l}  \cong  \left(\begin{array}{ccc}
                   1/2 & 1/2   &  0 \\
                     1/4 & 1/4  & 1/2 \\
                    1/4  &   1/4    & 1/2 \\
                    \end{array}\right),  \nonumber \\ 
 W_{i} & \cong & \left(\begin{array}{ccc}
                    0 & 1   &  0 \\
                     1/2 & 0  & 1/2 \\
                    1/2  &   0    & 1/2 \\
                    \end{array}\right), \hspace{.15in} 
   W_{h}  \cong  \left(\begin{array}{ccc}
                    0 & 1/2   &  1/2 \\
                     1/2 & 1/4  & 1/4 \\
                    1/2  &   1/4    & 1/4 \\
                    \end{array}\right),  \nonumber \\ 
   W_{d}  & \cong & \left(\begin{array}{ccc}
                    0 & 0   &  1 \\
                    1/2 & 1/2  & 0 \\
                    1/2  &  1/2   & 0 \\
                    \end{array}\right).
\end{eqnarray}
As a group, these matrices exhibit the remarkable simplicity of the 
PMNS matrix as $A$ varies from $0$ to $\infty$.  
Note that all of the matrices have 
at least one zero, $W_{1I}=0$, implying $x_{I}=y_{I}=0$.

\begin{figure}[ttt]
\caption{The numerical (solid) and approximate (dot-dashed) solutions
for $x_{1}(A)$, $x_{2}(A)$, and $x_{3}(A)$.  Note that $y_{i}(A) \simeq -x_{i}(A)$.} 
\centerline{\epsfig{file=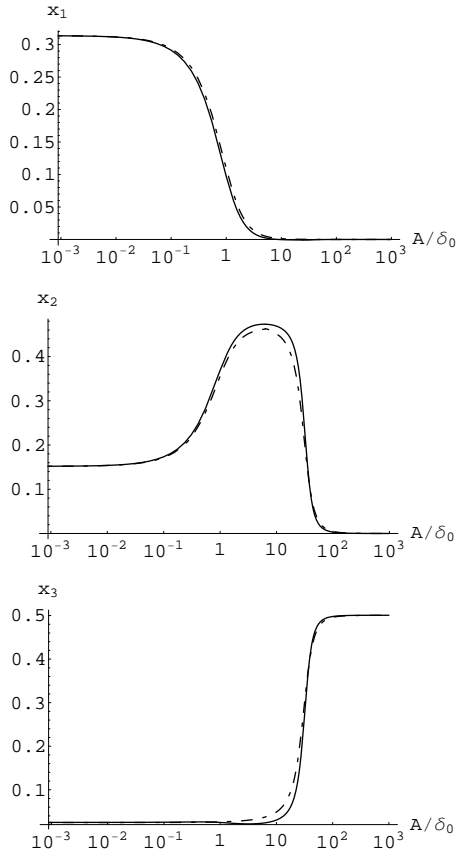,width=6 cm}}
\end{figure} 

The other feature, as mentioned before, is that they have equal elements
in their second and third rows, $W_{2i}=W_{3i}$.  This means that the $W$ matrix is
completely fixed by its first row, $W_{1i}$.
These elements, in turn, control $dD_{i}/dA$, Eq.~(\ref{eq:di}).
Thus, the progression of $W$ as a function of $A$ can be read off from the plot 
of $D_{i}(A)$, which is given in Fig. 1. 
Generally, the accuracy of the entries in Eq.~(\ref{eq:sum}) is of the order $10^{-2}$.
Note also that $A_{i}$ covers a rather large range, 
$2\delta_{0} \lesssim A_{i} \lesssim \Delta_{0}(1-2\epsilon^{2})$.
The validity of Eq.~(\ref{eq:sum}) will be confirmed by numerical integrations,
to be given in the next section.  Where applicable, they also agree with numerical results 
in the literature \cite{group}, after the proper change of variables is carried out.

\begin{figure}[ttt]
\caption{The mixing parameters in matter under the normal hierarchy (left column)
and the inverted hierarchy (right column), for both the $\nu$ sector (solid, $x_{i}$)
and the $\bar{\nu}$ sector (dashed, $\bar{x}_{i}$).  
Note that $x_{i}+y_{i} \simeq 0$ ($\bar{x}_{i}+\bar{y}_{i} \simeq 0$) for all $A$, 
and $x_{i}=\bar{x}_{i}$ in vacuum.} 
\centerline{\epsfig{file=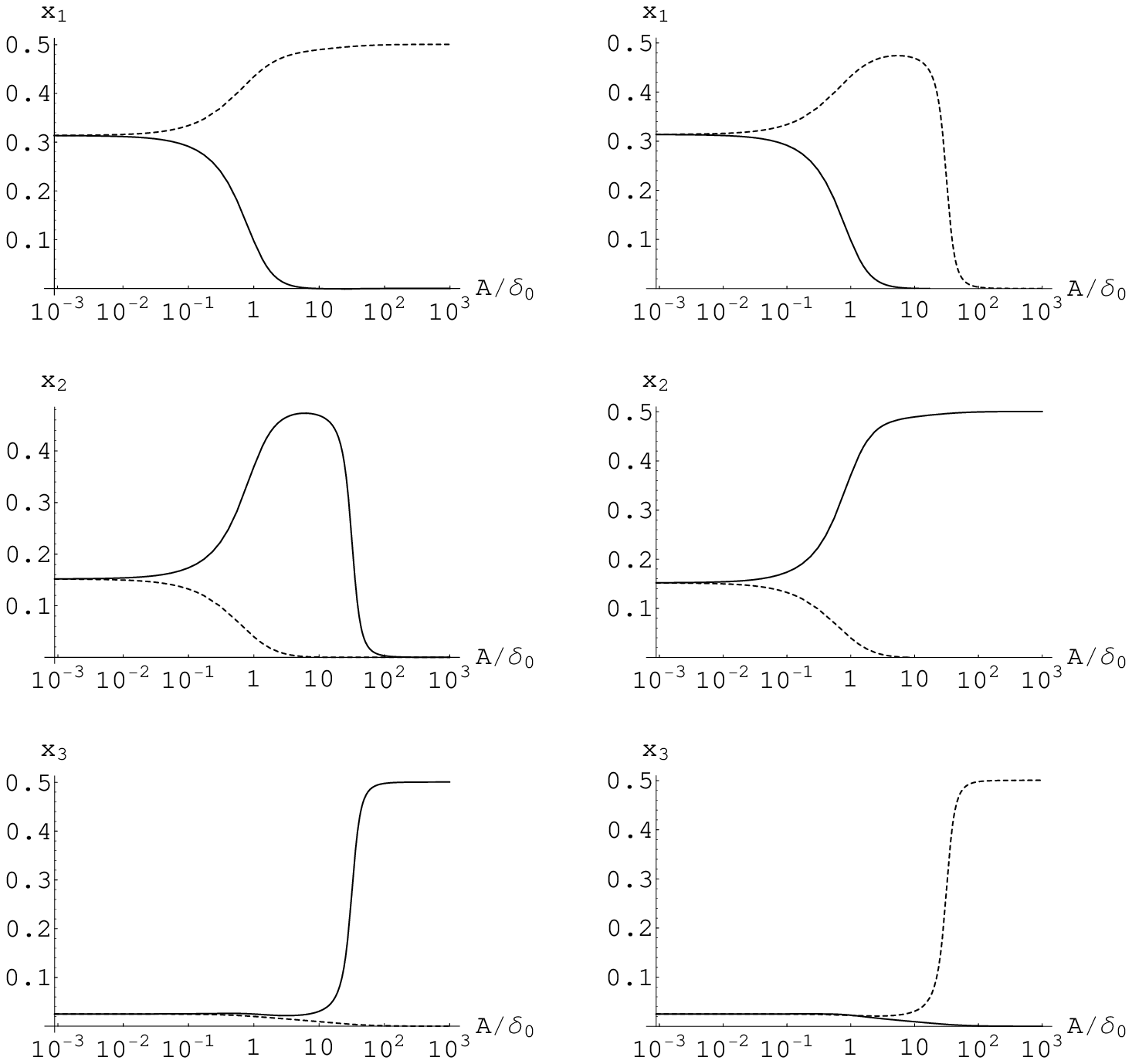,width=12 cm}}
\end{figure} 

\section{Numerical solutions}

It is straightforward to numerically integrate the evolution
equations for $(x,y)$.  
To do this we first obtain the vacuum expressions for the
$(x,y)$ parameters from Eqs. (6) and (8):
\begin{eqnarray}\label{eq:initial}
x_{10} & = & \frac{1}{6}(2-3\beta-2\epsilon^{2}), \hspace{.2in}
y_{10} =  \frac{1}{6}(-2-3\beta+2\epsilon^{2}), \nonumber \\
x_{20} & = & \frac{1}{6}(1-3\beta-\epsilon^{2}), \hspace{.2in}
y_{20}  =  \frac{1}{6}(-1-3\beta+\epsilon^{2}), \nonumber \\
x_{30} & = & \frac{1}{2} (\beta+\epsilon^{2}), \hspace{.2in}
y_{30}  =  \frac{1}{2}(\beta-\epsilon^{2}),
\end{eqnarray}
where $\xi=\eta=0$ is chosen and the terms in $\mathcal{O}(\beta \epsilon^{2})$
are ignored.  In addition, we choose the initial values 
$\epsilon=0.17$, $\beta =0.02$,
corresponding to the experimental bounds
$|V_{e3}|^{2} \leq 0.03$ \cite{data} and an assumed CP violation
phase $\cos \phi=1/4$.  With the input hierarchy $\delta_{0}/\Delta_{0}=1/32$,
the numerical results for $D_{i}$ and that for $(x_{i},y_{i})$ are then compared with the
approximate solutions obtained earlier (Eqs.~(\ref{low}),~(\ref{high})) in Fig. 1
and Fig. 2, respectively. The agreements are quite good.  

\begin{figure}[ttt]
\caption{The evolution of $J^{2}=x_{1}x_{2}x_{3}-y_{1}y_{2}y_{3}$ and
$\bar{J}^{2}=\bar{x}_{1}\bar{x}_{2}\bar{x}_{3}-\bar{y}_{1}\bar{y}_{2}\bar{y}_{3}$,
obtained from the numerical solutions for $\nu$ (solid)
and $\bar{\nu}$ (dashed) sectors.} 
\centerline{\epsfig{file=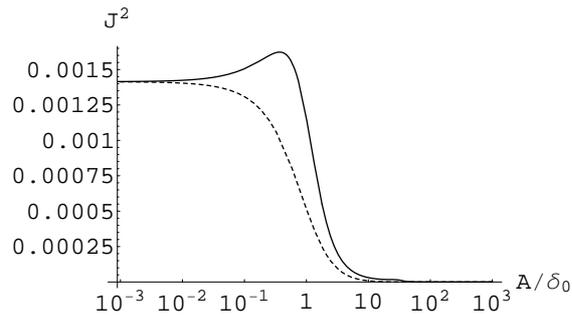,width=7.5 cm}}
\end{figure} 

\begin{figure}[ttt]
\caption{$W_{ij}$ for both the $\nu$ (solid lines) and $\bar{\nu}$ (dashed lines) sectors
are plotted as functions of $A/\delta_{0}$.  
Note that $W_{2i}=W_{3i}$,
and only the first two rows of $W_{ij}$ are shown in the figure.
The patterns of $W_{ij}$ can be readily deduced from Fig. 3.} 
\centerline{\epsfig{file=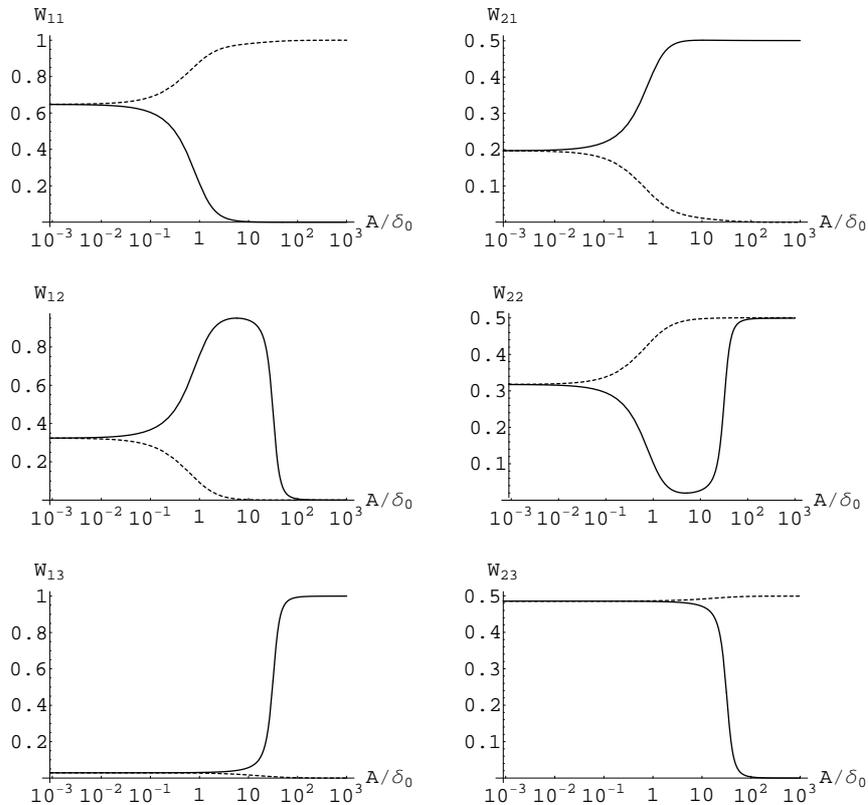,width=12 cm}}
\end{figure} 

\begin{figure}[ttt]
\caption{The mixing angles for the $\nu$ (left) and the $\bar{\nu}$ (right)
sectors in the standard parametrization: 
$\sin^{2}\theta_{12}$ (solid),
$\sin^{2}\theta_{23}$ (dot-dashed), and $\sin^{2}\theta_{13}$ (dashed), 
are plotted using the $(x,y)$ values in Fig. 3.
Note that $\theta_{23}$ for the $\nu$ sector remains constant 
($\theta_{23}=\pi/4$) for $A<A_{h}$ (higher resonance)
and deviates slightly from $\pi/4$ for $A>A_{h}$.
} 
\centerline{\epsfig{file=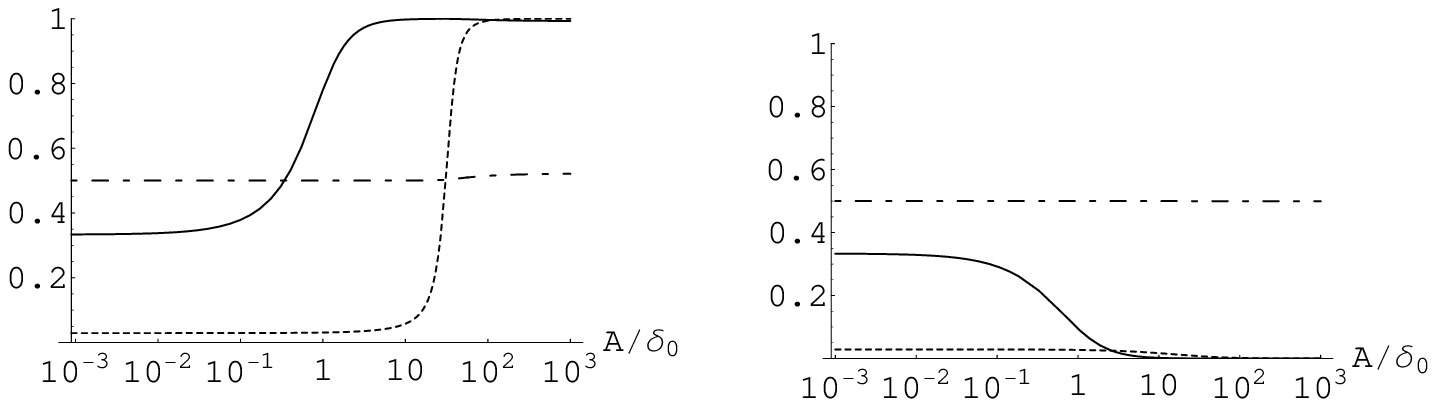,width=13 cm}}
\end{figure} 

Given the possible normal or inverted mass hierarchies, 
Fig. 3 summaries the evolution of the mixing parameters for both the $\nu$ and the $\bar{\nu}$
sectors.  Note that the parameters $(\bar{x}_{i},\bar{y}_{i})$ 
for the $\bar{\nu}$ sector can be obtained 
by replacing $A$ with $-A$ and $V$ with $V^{*}$ in that for the $\nu$ sector.
With the establishment of these ``basic solutions" for the mixing parameters,
we may easily study other physical quantities that are relevant to practical
calculations for the neutrino propagation in matter.
For illustration purpose, we plot some of the quantities numerically 
under the normal hierarchy in the following. The corresponding solutions
under the inverted hierarchy can be manipulated likewise.

The evolution of $J^{2}=x_{1}x_{2}x_{3}-y_{1}y_{2}y_{3}$ 
($\bar{J}^{2}=\bar{x}_{1}\bar{x}_{2}\bar{x}_{3}-\bar{y}_{1}\bar{y}_{2}\bar{y}_{3}$)
in matter is shown in Fig. 4.  
Compared to its vacuum value,
it is seen that, except for some enhancement for $J^{2}$ near $A=A_{l}$ and $A=A_{h}$,
the general trend is for it to decrease with $A$.
We note that near $A_{l}$, in the $1/(D_{1}-D_{2})$ dominance approximation,
$J^{2}(D_{1}-D_{2})^{2} \approx \mbox{constant}$, so while $(D_{2}-D_{1})$ goes through
a dip, $J^{2}$ has a bump near $A_{l}$, 
after which $J^{2}/J^{2}_{0} \approx (\delta_{0}/A)^{2}$, for $\delta_{0} \ll A \lesssim \Delta_{0}$.
A similar behaviour occurs near $A_{h}$, with $J^{2}(D_{2}-D_{3})^{2} \approx \mbox{constant}$,
although the effects are hardly noticeable. Similar numerical results were
also reached by solving directly the eigenvalue problem \cite{group}.
In addition,        
Fig. 5 shows the evolution of the first two rows of $W_{ij}$ in matter.  
We do not present the plots of $W_{3i}$ since they are almost
indistinguishable from those of $W_{2i}$.
The patterns of $W_{ij}$ in Eq.~(\ref{eq:sum}) are clearly seen from the plots.
Furthermore, the
mixing angles of the standard parametrization:
$\sin^{2}\theta_{12}$, $\sin^{2}\theta_{23}$, and $\sin^{2}\theta_{13}$,
are related to the $(x,y)$ parameters:
\begin{equation}
\sin^{2}\theta_{12}=1/(1+\frac{x_{1}-y_{1}}{x_{2}-y_{2}}),
\end{equation}
\begin{equation}
\sin^{2}\theta_{23}=1/(1+\frac{x_{1}-y_{2}}{x_{2}-y_{1}}),
\end{equation}
\begin{equation}
\sin^{2}\theta_{13}=x_{3}-y_{3}.
\end{equation}
The numerical results for both the $\nu$ and $\bar{\nu}$ sectors are shown in Fig. 6.
These plots are in agreement with those in the literature \cite{group}.
For $\theta_{12}$ and $\theta_{13}$, note the characteristic step-function
resonance behaviors near $A_{l}$ and $A_{h}$.  Also, $\theta_{23} \cong \pi/4$
is a reflection of $W_{23} \cong W_{33}$, for all $A$.  The phase angle $(\phi)$
is not plotted since it also remains constant due to the invariance
of $S_{\phi}\sin2\theta_{23}$.


\section{Conclusion}
 
Understanding the propagation of neutrinos through matter is one of
the core problems in neutrino physics.  In a medium of constant density,
it is well-known that the electron neutrino acquires an induced mass which
alters both the eigenvalues and the mixing matrix of the neutrinos.
Traditionally, one studies directly the eigenvalue problem of the effective Hamiltonian.
The neutrino parameters are expressed as complicated formulas in terms
of the induced mass and their values in vacuum.
One then resorts to numerical plots by assuming specific values for these
partially known parameters.  The drawback of this method is the lack of
insights into the nature of the solutions, and it is not easy to gain an
overview of the mixing as a function of the induced mass.

In this paper we try a different approach, by finding the evolution equations
of the neutrino parameters as a function of the induced mass.
The resulting equations, when written in terms of a rephasing invariant parametrization,
turn out to be manageable and we are able to find simple, approximate, solutions
with the help of two important features of the vacuum neutrino parameters.
1) The two measured mass differences are widely separated so that the two-flavor
resonance approximation becomes applicable.  2) The vacuum PMNS matrix has an
approximate $\mu-\tau$ symmetry, which is preserved by the set of evolution equations.
The result is summarized in Eq.~(\ref{eq:sum}), showing the striking simplicity
of the neutrino mixing matrix as a function of $A$.  
Approximate solutions for the parameters are explicitly given in 
Eqs.~(\ref{low}) and ~(\ref{high}).  
The evolution equations
also facilitate the derivation of ``matter invariants", given in Eq.~(\ref{JD})
and ~(\ref{eq:in2}).  In addition, there are also ``partial matter invariants",
Eqs.~(\ref{pmiL}) and ~(\ref{pmiH}).  
These are useful in obtaining properties of the various parameters without
performing detailed calculations.

Based on the incomplete measurements that exist for the vacuum parameters,
our analyses show that those in matter, to a good approximation, can already
be determined.  We hope that these results will be helpful in the
exploration of the physics of neutrino propagation in matter.

\acknowledgments 
S.H.C. is supported by the National 
Science Council of Taiwan, grant No. NSC 98-2112-M-182-001-MY2.

\appendix

\section{Neutrino transition probabilities in $(x,y)$ parameters}

As the neutrinos travel through a baseline $L$ in matter of constant density,
the flavor transition probability is given by
\begin{eqnarray}
P(\nu_{\alpha} \rightarrow \nu_{\beta})=\delta_{\alpha \beta}&-&
4\sum_{j>i}Re(V_{\alpha i}V^{*}_{\beta j}V^{*}_{\alpha j}V_{\beta j})
\sin^{2}(D_{ij}) \nonumber \\
 &+& 2\sum_{j>i}Im(V_{\alpha i}V^{*}_{\beta i}V^{*}_{\alpha j}V_{\beta j})
\sin(2D_{ij}),
\end{eqnarray}
where $D_{ij} \equiv (D_{i}-D_{j})L/4E$.
For $\alpha \neq \beta$, we obtain the explicit expression,
\begin{eqnarray}\label{eq:im}
P(\nu_{\alpha} \rightarrow \nu_{\beta})=&-&4[Re(V_{\alpha 1}V^{*}_{\beta 1}
V^{*}_{\alpha 2}V_{\beta 2}) \sin^{2}(D_{12})+Re(V_{\alpha 1}V^{*}_{\beta 1}
V^{*}_{\alpha 3}V_{\beta 3}) \sin^{2}(D_{13}) \nonumber \\
&+&Re(V_{\alpha 2}V^{*}_{\beta 2}
V^{*}_{\alpha 3}V_{\beta 3}) \sin^{2}(D_{23})]  \nonumber \\
   &  +  & 2[Im(V_{\alpha 1}V^{*}_{\beta 1}
V^{*}_{\alpha 2}V_{\beta 2}) \sin(2D_{12})+Im(V_{\alpha 1}V^{*}_{\beta 1}
V^{*}_{\alpha 3}V_{\beta 3}) \sin(2D_{13}) \nonumber \\
&+&Im(V_{\alpha 2}V^{*}_{\beta 2}
V^{*}_{\alpha 3}V_{\beta 3}) \sin(2D_{23})].
\end{eqnarray}
For a specific process, $e.g.$, $P(\nu_{\mu} \rightarrow \nu_{e})$, we have
\begin{eqnarray}
Re(V_{21}V_{12}V^{*}_{22}V^{*}_{11})&=&x_{2}x_{3}+x_{1}y_{2}-y_{1}y_{2}-y_{2}y_{3} \equiv F^{\mu e}_{21}, \nonumber \\
Re(V_{21}V_{13}V^{*}_{23}V^{*}_{11})&=&-x_{1}x_{3}-x_{2}x_{3}+x_{3}y_{1}+y_{2}y_{3} \equiv F^{\mu e}_{31}, \nonumber \\
Re(V_{22}V_{13}V^{*}_{23}V^{*}_{12})&=&x_{1}x_{3}+x_{2}y_{3}-y_{1}y_{3}-y_{2}y_{3} \equiv F^{\mu e}_{32},
\end{eqnarray}
and the probability,
\begin{eqnarray}
P(\nu_{\mu} \rightarrow \nu_{e})=&-&4[F^{\mu e}_{21} \sin^{2}(D_{21})+ 
F^{\mu e}_{31} \sin^{2}(D_{31})+F^{\mu e}_{32} \sin^{2}(D_{32})] \nonumber \\
  &  +  & 8J\sin(D_{21})\sin(D_{31})\sin(D_{32}),
  \end{eqnarray}
where $Im[V_{\alpha i}V_{\beta j}V^{*}_{\alpha j}V^{*}_{\beta i}]
=J\sum_{\gamma,k}\epsilon_{\alpha\beta\gamma}\epsilon_{ijk}$
has been used in reducing the sum of the imaginary parts in Eq.~(\ref{eq:im}).
In addition, the probability for the T-conjugate process takes the form
\begin{eqnarray}
P(\nu_{e} \rightarrow \nu_{\mu})=&-&4[F^{e \mu}_{21} \sin^{2}(D_{21})+ 
F^{e \mu}_{31} \sin^{2}(D_{31})+F^{e \mu}_{32} \sin^{2}(D_{32})] \nonumber \\
  &  -  & 8J\sin(D_{21})\sin(D_{31})\sin(D_{32}),
  \end{eqnarray}
where
\begin{eqnarray}
F^{e \mu}_{21} & = &-x_{1}x_{2}-x_{1}x_{3}+x_{1}y_{2}+y_{1}y_{3}, \nonumber \\
F^{e \mu}_{31} & = &x_{1}x_{2}+x_{3}y_{1}-y_{1}y_{2}-y_{1}y_{3}, \nonumber \\
F^{e \mu}_{32} & = &-x_{1}x_{2}-x_{2}x_{3}+x_{2}y_{3}+y_{1}y_{2}.
\end{eqnarray}
We may verify the relation $F^{\mu e}_{ij}=F^{e \mu}_{ij}$ using Eq.(4).

The explicit probabilities for other processes can be derived following the same procedure:
\begin{eqnarray}
P(\nu_{e} \rightarrow \nu_{\tau})=&-&4[(x_{1}x_{3}+x_{2}y_{1}-y_{1}y_{2}-y_{1}y_{3})\sin^{2}(D_{21}) \nonumber \\
                    & + & (-x_{1}x_{2}-x_{1}x_{3}+x_{1}y_{3}+y_{1}y_{2})\sin^{2}(D_{31}) \nonumber \\
                    & + & (x_{1}x_{2}+x_{3}y_{2}-y_{1}y_{2}-y_{2}y_{3})\sin^{2}(D_{32})] \nonumber \\
  & + & 8J\sin(D_{21})\sin(D_{31})\sin(D_{32})
  \end{eqnarray}
  
  \begin{eqnarray}
P(\nu_{\tau} \rightarrow \nu_{e})=&-&4[(-x_{1}x_{2}-x_{2}x_{3}+x_{2}y_{1}+y_{2}y_{3})\sin^{2}(D_{21}) \nonumber \\
                    & + & (x_{2}x_{3}+x_{1}y_{3}-y_{1}y_{3}-y_{2}y_{3})\sin^{2}(D_{31}) \nonumber \\
                    & + & (-x_{1}x_{3}-x_{2}x_{3}+x_{3}y_{2}+y_{1}y_{3})\sin^{2}(D_{32})] \nonumber \\
  & - & 8J\sin(D_{21})\sin(D_{31})\sin(D_{32})
  \end{eqnarray}

 \begin{eqnarray} 
  P(\nu_{\mu} \rightarrow \nu_{\tau})=&-&4[(-x_{1}x_{3}-x_{2}x_{3}+x_{3}y_{3}+y_{1}y_{2})\sin^{2}(D_{21}) \nonumber \\
                    & + & (x_{1}x_{3}+x_{2}y_{2}-y_{1}y_{2}-y_{2}y_{3})\sin^{2}(D_{31}) \nonumber \\
                    & + & (-x_{1}x_{2}-x_{1}x_{3}+x_{1}y_{1}+y_{2}y_{3})\sin^{2}(D_{32})] \nonumber \\
  & - & 8J\sin(D_{21})\sin(D_{31})\sin(D_{32})
  \end{eqnarray}
  
  \begin{eqnarray}
P(\nu_{\tau} \rightarrow \nu_{\mu})=&-&4[(x_{1}x_{2}+x_{3}y_{3}-y_{1}y_{3}-y_{2}y_{3})\sin^{2}(D_{21}) \nonumber \\
                    & + & (-x_{1}x_{2}-x_{2}x_{3}+x_{2}y_{2}+y_{1}y_{3})\sin^{2}(D_{31}) \nonumber \\
                    & + & (x_{2}x_{3}+x_{1}y_{1}-y_{1}y_{2}-y_{1}y_{3})\sin^{2}(D_{32})] \nonumber \\
  & + & 8J\sin(D_{21})\sin(D_{31})\sin(D_{32})
  \end{eqnarray}
We may also write down the expressions 
for the $\bar{\nu}$ sector, $P(\bar{\nu}_{\alpha} \rightarrow \bar{\nu}_{\beta})$,
by replacing the parameters for the $\nu$ sector 
with that for the $\bar{\nu}$ sector:
$x \rightarrow \bar{x}$, $y \rightarrow \bar{y}$, $D_{ij} \rightarrow \bar{D}_{ij}$,
and thus $F^{\alpha \beta}_{ij} \rightarrow \bar{F}^{\alpha \beta}_{ij}$, 
$J \rightarrow \bar{J}$.  As an example, 
the probability $P(\bar{\nu}_{\mu} \rightarrow \bar{\nu}_{e})$ is given by
\begin{eqnarray}
P(\bar{\nu}_{\mu} \rightarrow \bar{\nu}_{e})=&-&4[\bar{F}^{\mu e}_{21} \sin^{2}(\bar{D}_{21})+ 
\bar{F}^{\mu e}_{31} \sin^{2}(\bar{D}_{31})+\bar{F}^{\mu e}_{32} \sin^{2}(\bar{D}_{32})] \nonumber \\
  &  -  & 8\bar{J}\sin(\bar{D}_{21})\sin(\bar{D}_{31})\sin(\bar{D}_{32}),
  \end{eqnarray}
where
\begin{eqnarray}
\bar{F}^{\mu e}_{21} & = &\bar{x}_{2}\bar{x}_{3}+\bar{x}_{1}\bar{y}_{2}-\bar{y}_{1}\bar{y}_{2}-\bar{y}_{2}\bar{y}_{3}, \nonumber \\
\bar{F}^{\mu e}_{31} & = &-\bar{x}_{1}\bar{x}_{3}-\bar{x}_{2}\bar{x}_{3}+\bar{x}_{3}\bar{y}_{1}+\bar{y}_{2}\bar{y}_{3}, \nonumber \\
\bar{F}^{\mu e}_{32} & = &\bar{x}_{1}\bar{x}_{3}+\bar{x}_{2}\bar{y}_{3}-\bar{y}_{1}\bar{y}_{3}-\bar{y}_{2}\bar{y}_{3}.
\end{eqnarray}
The evolution equations and the analytic, approximate, solutions for $(\bar{x}_{i},\bar{y}_{i})$ 
can be obtained following the same method outlined in this work.
The numerical solutions for $(\bar{x}_{i},\bar{y}_{i})$ are shown in Fig. 3. 
Note that 1) $F^{\alpha \beta}_{ij}=\bar{F}^{\alpha \beta}_{ij}=
F^{\beta \alpha}_{ij}=\bar{F}^{\beta \alpha}_{ij}$ in vacuum; and
2) the functions $F^{\alpha \beta}_{ij}$, $F^{\beta \alpha}_{ij}$,  
$\bar{F}^{\alpha \beta}_{ij}$, and $\bar{F}^{\beta \alpha}_{ij}$ can take
varied forms in terms of $(x,y)$ and $(\bar{x},\bar{y})$
since there are different ways \cite{Kuo:05} of reducing 
$Re(V_{\alpha i}V^{*}_{\beta j}V^{*}_{\alpha j}V_{\beta j})$.


\begin{thebibliography}{}
\bibitem{Wij}
Clearly, if $W_{2i}=W_{3i}$, $w_{1i}=0$, 
and $x_{i}+y_{i}=0$.
Conversely, given, $e.g.$, $W_{22}W_{33}-W_{23}W_{32}=0$ and $W_{21}W_{33}-W_{23}W_{31}=0$,
using unitarity, one finds $W_{2i}=W_{3i}$.
\end{thebibliography}
\end{document}